\documentclass[aps,twocolumn,groupedaddress]{revtex4}
\usepackage[dvips]{color}
\usepackage{graphicx,color}
\usepackage{bm}        % for math
\usepackage{amssymb}   % for math
\usepackage{amsmath}
\usepackage{cases}

\begin{document}

\title{Probing Majorana Flat Bands in Nodal $d_{x^2-y^2}$-wave Superconductors with Rashba Spin-Orbit Coupling}

\author{Noah F. Q. Yuan, Chris L.M. Wong and K. T. Law $^\dagger$ } 

\affiliation{Department of Physics, Hong Kong University of Science and Technology \\ Clear Water Bay, Kowloon, Hong Kong, China}

\begin{abstract}
We show that Majorana fermions associated with Majorana flat bands emerge as zero energy modes on the [110] edge of single layer or multilayer nodal superconductors with $d_{x^2-y^2}$-wave pairing and Rashba spin-orbit coupling. Moreover, as long as the global inversion symmetry of the single layer or multilayer superconductor is broken, and in the presence of an in-plane magnetic field parallel to the [110] edge, the Majorana fermions together with usual fermionic Andreev bound states induce a triple-peak structure in tunnelling spectroscopy experiments. Importantly, we show that the zero bias conductance peak is induced by Majorana fermions. Therefore, tunnelling spectroscopy can be used to probe Majorana fermions in nodal $d_{x^2-y^2}$-wave superconductors.
\end{abstract}

\maketitle

\section{Introduction}
Majorana fermions, \cite{Wilczek, Kitaev} which act as their own anti-particles and emerge as zero energy excitations in topological superconductors, have been the subject of intense theoretical \cite{Kitaev,RG,FK, KH,Beenakker, Alicea, Franz,ST} and experimental studies \cite{Kou, Deng, Das, Marcus}. Majorana fermions are topologically protected in the sense that they cannot be removed by perturbations to the superconductor unless certain symmetries are broken or the bulk gap of the superconductor is closed.

Remarkably, recent development shows that Majorana fermions exist in nodal topological superconductors with gapless bulk spectra \cite{Beri, Sato1, Sato2, Sato3, Schnyder1, Schnyder2, Balents, Wang, Wong}. For example, Majorana edge modes, which are associated with Majorana flat bands, can be found in 2D nodal $d_{xy} +p$-wave superconductors \cite{Sato1,Sato2,Sato3}. It is also shown that zero energy Majorana flat bands can appear on the surface of 3D non-centrosymmetric superconductors which have topologically stable line nodes in the bulk \cite{Schnyder1,Schnyder2}. Majorana edge modes, which are robust against disorder, can also be created by tuning a fully gapped $p \pm ip $-wave superconductor into the nodal regime by an external magnetic field \cite{Wong}.

It has been shown that $d_{x^2-y^2}$-wave superconductors without Rashba spin-orbit coupling (SOC) support zero energy edge states on the [110] edge (or equivalent edges) \cite{Hu}, which are associated with fermionic flat bands of the bulk energy spectrum \cite{Hatsugai}. Recently, it has been shown that in the presence of Rashba SOC, Majorana zero modes associated with Majorana flat bands coexist with fermionic zero energy modes on the [110] edge \cite{Sato1,Wang}, as depicted schematically in Fig.1a. In this work, we first show that for the single layer case, and in the presence of an in-plane magnetic field parallel to the [110] edge, the Majorana zero energy modes stay at zero energy while the zero energy fermionic modes are lifted to finite energy. This results in a triple-peak structure in the Andreev reflection type tunnelling experiments. We show that the conductance peak arising at zero voltage bias is due to the presence of Majorana fermions. This is different from the double-peak case in the usual $d_{x^2-y^2}$-wave superconductor without Majorana fermions \cite{Tanaka}.

\begin{figure}
\centering
\includegraphics[width=3.2in]{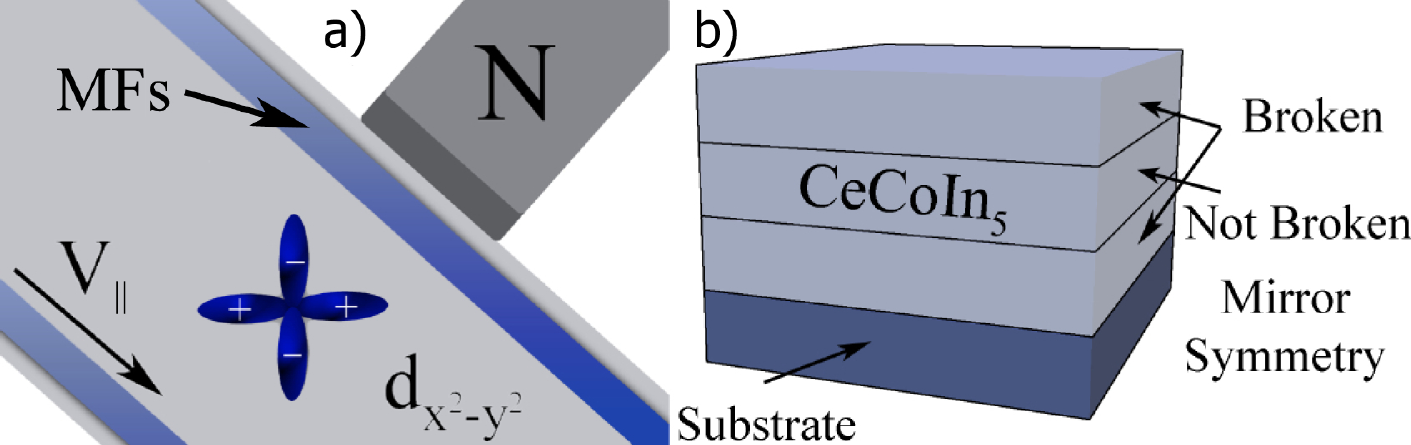}
\caption{\label{Fig1} a) A schematic picture of $d_{x^2-y^2}$-wave superconductor with Rashba SOC. Zero energy Majorana fermions (MFs) and zero energy fermions coexist on the [110] edge in the absence of a magnetic field. An in-plane magnetic field along the edge direction does not lift the zero energy Majorana modes to finite energy but can lift the fermionic modes to finite energy. A normal lead is attached to the [110] edge. b) A thin film of CeCoIn$_{5}$, a candidate of multilayer $d_{x^2-y^2}$-wave superconductors with Rashba SOC due to the local inversion symmetry breaking. The Rashba SOC can be different in different layers as the local inversion symmetry is broken differently.}
\end{figure}

Single layer $d_{x^2-y^2}$-wave superconductors with Rashba SOC are yet to be identified experimentally. In the experimentally more relevant case of a multilayer structure as depicted in Fig.1b, we show that: 1) in the absence of a magnetic field, the Majorana zero modes can still survive even when multiple topologically non-trivial layers are coupled to each other. The Majorana fermions do not hybridize with each other. This is in sharp contrast to the Majorana fermions in time-reversal symmetry breaking topological superconductors in which coupling two Majorana fermions lifts the Majorana fermions to finite energy \cite{Potter}. This is due to the fact that the 2D Hamiltonian of the multilayer system can be reduced to a sum of 1D Hamiltonians in the BDI class, which are classified by integers \cite{SRFL,Teo}. 2) In the presence of an in-plane magnetic field parallel to the [110] edge, the 1D Hamiltonians are reduced to D class, which are classified by $Z_{2}$ numbers \cite{SRFL,Teo}. We show that the Majorana fermion zero modes can survive only when the global inversion symmetry of the multilayer system is broken. Furthermore, the Majorana zero modes in the multilayer case can also induce zero bias conductance peaks (ZBCPs), which cannot be removed by an in-plane magnetic field. Therefore, tunnelling spectroscopy can be used to probe Majorana fermions in nodal topological superconductors.

%In the following sections, we first study a single-layer $d_{x^2-y^2}$-wave superconductor with Rashba spin-orbit coupling and show that Majorana fermions emerge on the [110] edge which are associated with Majorana flat bands. Second, we show that the Majorana flat band are robust against in-plane magnetic field due to extra symmetry of the Hamiltonian. Third, the tunnelling conductance from a normal lead to the [110] edge is studied as depicted in Fig.1a. Finally, we study a multilayer $d_{x^2-y^2}$-wave superconductor with inhomogeneous Rashba spin-orbit coupling which can be relevant to thin films of CeCIn$_{5}$ with $d_{x^2-y^2}$-wave pairing and strong spin-orbit coupling.

\section{BDI classification of the single layer cases}
It has been pointed out recently that Majorana flat bands can emerge in a single layer $d_{x^2-y^2}$-wave superconductor with Rashba SOC \cite{Sato1, Wang}. In this section, we show that the 2D Hamiltonian can be reduced to a sum of independent 1D Hamiltonians in the BDI class which are classified by integer numbers. This classification is important for the stability of Majorana fermions in the multilayer case.

To study the Majorana fermions of a $d_{x^2-y^2}$-wave superconductor on the [110] edge, we denote the momentum parallel and perpendicular to the [110] direction as $k_{\parallel}$ and $k_{\perp}$ respectively and the Hamiltonian in momentum space is:
\begin{equation}
H_{l}(\bm k)=
\begin{pmatrix}
H_{l 0}(\bm k)&\Delta(\bm k)\\
\Delta^{\dagger}(\bm k)&-H_{l0}^{*}(-\bm k)
\end{pmatrix}
.
\end{equation}
Here, $l$ is the layer index which is relevant in the multilayer case. Only the single layer case is considered in this section. The basis is $ \psi_{l \bm k }=(c_{l \bm k \uparrow},c_{l \bm k \downarrow},c_{l -\bm k \uparrow}^{\dagger},c_{l -\bm k \downarrow}^{\dagger}) $ where $c_{l \bm k\alpha}$ is an electron operator on layer $l$ with momentum $\bm k$ and spin $\alpha$. $ \Delta(\bm k)=\Delta_{d} \sin k_{\parallel}\sin  k_{\perp} i \sigma_y $ is the pairing matrix and $\Delta_{d}$ is the pairing strength. The normal part of the Hamiltonian is $ H_{l 0}(\bm k)=(-2t\cos k_{\parallel}-2t\cos k_{\perp}-\mu)\sigma_{0}+[\bm g_{l}(\bm k)+\bm V]\cdot\bm\sigma $ where $t $ is the intralayer hopping amplitude, $ \bm g_{l}(\bm k)\cdot\bm\sigma $ describes the Rashba SOC and $ \bm V $ is the magnetic field. Here, $ \bm g_{l}(\bm k)=\alpha_{l}(\sin k_{\parallel},-\sin k_{\perp},0)$ is chosen and $\alpha_{l}$ is the Rashba strength on layer $l$.

A tight-binding model on a square lattice which reproduces $H_{l}(\bm k)$ in momentum space, can be written as:
\begin{equation}
\begin{array}{ll}
H_{l,tb} = & H_{l,t} + H_{l,\text{SO}}+H_{l,\text{SC}} +H_{l,{\bm V}}, \\
H_{l,t}= & \sum_{\mathbf{R},\mathbf{d}, \alpha} -t(c^{\dagger}_{l,\mathbf{R+d},\alpha}c_{l,\mathbf{R}, \alpha}+h.c.)-\mu c^{\dagger}_{l,\mathbf{R},\alpha}c_{l,\mathbf{R}, \alpha}\\
H_{l,SO}= & \sum_{\mathbf{R,d},\alpha,\beta}-\frac{i}{2} \alpha_{l} c^{\dagger}_{l,\mathbf{R+d},\alpha} \hat{\bm{z}}\cdot(\bm{\sigma}_{\alpha \beta}\times \mathbf{d})c_{l,\mathbf{R},\beta}+h.c.\\
H_{l,SC}= & \sum_{\mathbf{R}}- \frac{\Delta_{d}}{8} (c^{\dagger}_{l,\mathbf{R+d_{\parallel}+d_{\perp}},\uparrow}c^{\dagger}_{l,\mathbf{R},\downarrow}-c^{\dagger}_{l,\mathbf{R+d_{\parallel}+d_{\perp}},\downarrow}c^{\dagger}_{l,\mathbf{R},\uparrow} \\ 
&-c^{\dagger}_{l,\mathbf{R+d_{\parallel}-d_{\perp}},\uparrow}c^{\dagger}_{l,\mathbf{R},\downarrow}+c^{\dagger}_{l,\mathbf{R +d_{\parallel}-d_{\perp}},\downarrow}c^{\dagger}_{l,\mathbf{R},\uparrow}  + h.c. )\\
H_{l,{\bm V}}= & \sum_{\mathbf{R},\alpha,\beta} c^{\dagger}_{l,\mathbf{R},\alpha}{\bm{V}}\cdot\bm{\sigma}_{\alpha \beta}c_{l,\mathbf{R},\beta}.
\end{array}  \label{Hq1D}
\end{equation}

Here, ${\bm d}$ represents the vectors connecting the nearest neighbour sites, with $\bm d_{\parallel}$ ($\bm d_{\perp}$) connecting sites parallel (perpendicular) to the [110] edge. The magnetic field strength parallel to the [110] edge is denoted as $V_{\parallel}$. To study the energy spectra in the presence of an edge, we apply periodic boundary conditions in the direction parallel to the edge and open boundary conditions in the direction perpendicular to the edge. The energy spectra are shown in Fig.2.

It is well known that in the absence of Rashba SOC, a $ d_{x^2-y^2} $-wave superconductor is nodal and possesses zero energy fermionic states on the [110] edge \cite{Hu, Hatsugai}. As shown in Fig.2a, it is evident that in the absence of Rashba SOC, there are zero energy fermionic states for a wide range of $k_{\parallel}$. The flat band connects the nodal points of the bulk. It is important to note that all states are doubly degenerate for any given $k_{\parallel}$ due to time-reversal symmetry and inversion symmetry of the Hamiltonian.

In the presence of the Rashba SOC, the inversion symmetry is broken and, as shown in Fig.2b, each nodal point at finite $k_{\parallel}$ is split into two nodal points. Interestingly, there are flat bands which connect the split bulk nodal points. In the rest of this section, we show that each zero energy mode for a particular $k_{\parallel}$  of the flat band, highlighted in red in Fig.2b, is associated with a Majorana fermion localized on the [110] edge.

\begin{figure}
\centering
\includegraphics[width=3.2in]{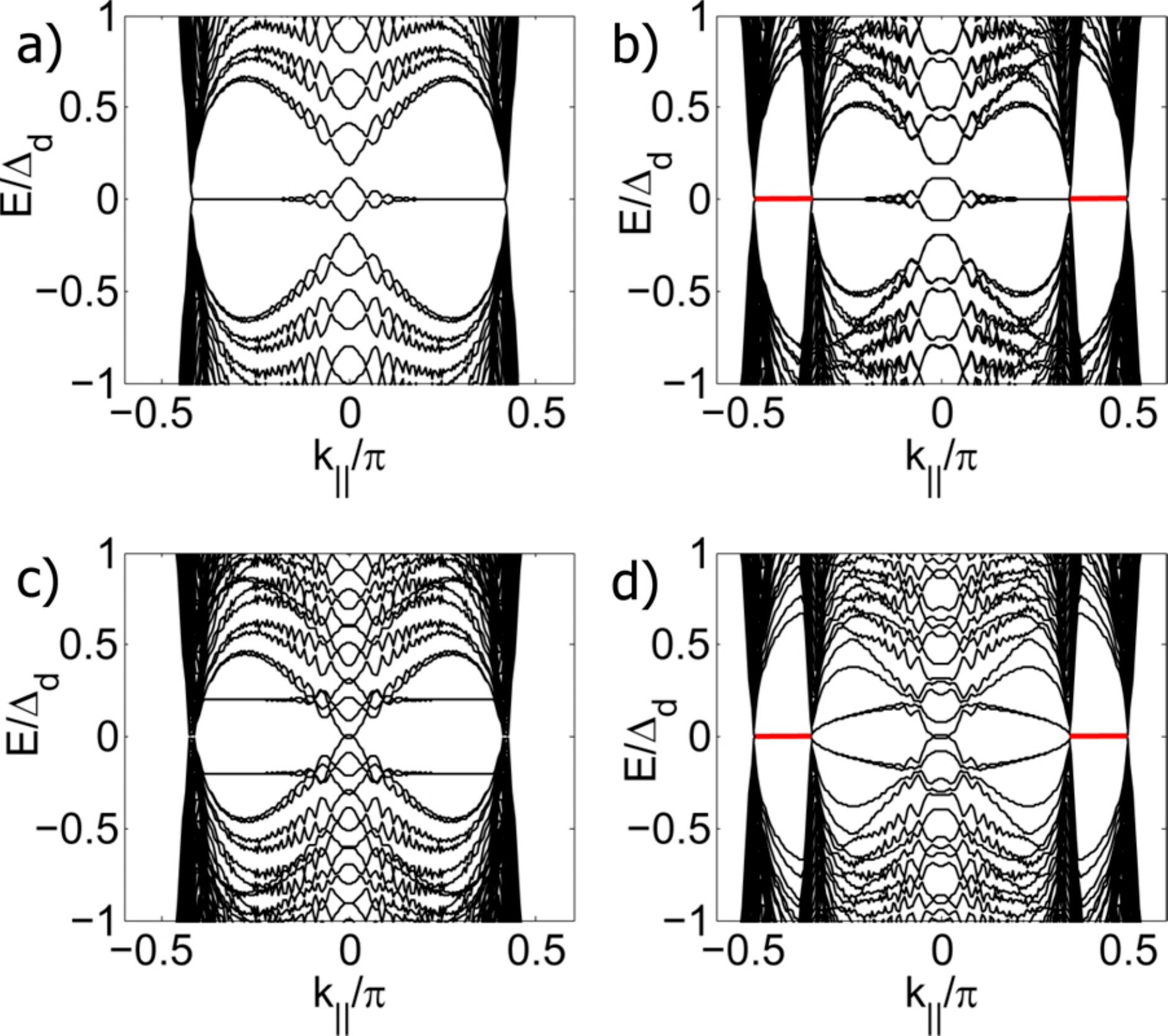}
\caption{\label{Fig2} The energy spectra of $d_{x^2-y^2}$-wave superconductors as functions of $k_{\parallel}$. Periodic boundary conditions in the [110] direction and open boundary conditions in the $[1\overline{1}0]$ direction are assumed. Along the $[1\overline{1}0]$ direction, there are 100 sites. The tight-binding parameters are: $t=10$, $\Delta_{d} =2$, $\mu=-25$ for all the figures. a) In the absence of Rashba SOC and magnetic field, there are no Majorana modes. b) In the presence of Rashba SOC, with $\alpha_{l}=2.5\Delta_{d}$ and ${\bm V}=\bm 0$, the Majorana flat bands, highlighted in red, emerge. c) $\alpha_{l}=0$, ${V}_{\parallel}=0.2\Delta_{d}$. The zero energy flat band is lifted to finite energy. d)  $\alpha_{l}=2.5 \Delta_{d}$ and ${{ V}_{\parallel}}=0.2\Delta_{d}$. The Majorana flat band is robust but the fermionic flat bands are lifted to finite energy by the magnetic field. The Majorana flat bands are highlighted in red.}
\end{figure}

To understand the origin of the Majorana flat bands, we note that the Hamiltonian $H_{l}(\bm k)$ respects the usual time-reversal symmetry $T= i \sigma_{y} \otimes \tau_{0} K $ and particle-hole symmetry $P= \sigma_{0} \otimes \tau_{x} K$, where $\sigma_{i}$ and $\tau_{i}$ are Pauli matrices which operate on the spin and the particle-hole space respectively. $K$ is the complex conjugate operator. Therefore, the Hamiltonian is in the DIII class according to the symmetry classification \cite{SRFL,Teo}. Unfortunately, since the bulk is nodal, the Hamiltonian cannot be classified by any topological invariant associated with DIII class. 

However, we note that the Hamiltonian respects a mirror symmetry $M=i \sigma_{y} \otimes\tau_{z}$ where $M H_{l} (k_{\parallel}, k_{\perp})M^{-1} = H_{l} (-k_{\parallel}, k_{\perp})$. Combining $M$ with $T$ and $P$, the Hamiltonian satisfies the time-reversal like symmetry $T_{1d}=MT= -\sigma_{0}\otimes \tau _{z}K $  where $T_{1d}H_{l}(k_{\parallel},k_{\perp})T_{1d}^{-1}=H_{l}(k_{\parallel},-k_{\perp})$ and the particle-hole like symmetry $P_{1d}=MP= -\sigma_{y} \otimes \tau_{y}K $ where $P_{1d}H_{l}(k_{\parallel},k_{\perp})P_{1d}^{-1}=-H_{l}(k_{\parallel},-k_{\perp})$. Since $P_{1d}$ and $T_{1d}$ do not operate on $k_{\parallel}$, $H_{l}(k_{\parallel},k_{\perp})$ can be regarded as a 1D Hamiltonian $H_{l,k_{\parallel}}(k_{\perp})$ where $k_{\parallel}$ is a tuning parameter. As $P_{1d}^{2}=T_{1d}^2=1$ and the Hamiltonian also satisfies the chiral symmetry $C=P_{1d}T_{1d}$ with $C H_{l,k_{\parallel}}(k_{\perp}) C^{-1}= -H_{l,k_{\parallel}}(k_{\perp})$, $H_{l,k_{\parallel}}(k_{\perp})$ with any specific $k_{\parallel}$ is in the BDI class \cite{SRFL,Teo} which is classified by an integer topological invariant number $N_{l,BDI}$. 

In the basis $U \psi_{l}$ which diagonalizes $P_{1d}T_{1d}$, the Hamiltonian can be off-diagonalized such that:
\begin{equation}
U H_{l,k_{\parallel}}(k_{\perp}){U}^{-1}=
\begin{pmatrix}
0&q_{l}\\
q^{\dagger}_{l}&0
\end{pmatrix}
\end{equation}
and the topological invariant is a winding number \cite{Sato3, Sau2, Wong2}
\begin{equation}
N_{l,BDI}(k_{\parallel})=\frac{-i}{\pi} \int_{k_{\perp}=0}^{k_{\perp}=\pi} \frac{dz_{l,k_{\parallel}}(k_{\perp})}{z_{l,k_{\parallel}}(k_{\perp})},  \label{NBDI1}
\end{equation}
where 
\begin{equation}
z_{l,k_{\parallel}}(k_{\perp})=\text{det}[q_{l}(k_{\parallel},k_{\perp})]/|\text{det}[q_{l}(k_{\parallel},k_{\perp})]|.  \label{NBDI2}
\end{equation}
%\begin{equation}
%N_{l, BDI}(k_{\parallel})=\frac{1}{2\pi }\text{Im}\int_{0}^{2\pi}dk_{\perp}\log\det q_{l}(k_{\parallel} ,k_{\perp}). \label{NBDI}
%\end{equation}
It can be shown that the topological invariant is:
\begin{numcases}{N_{l,BDI}(k_{\parallel})=}\nonumber
0& $k_{+} < |k_{\parallel}|<\pi $\\ \nonumber
\text{sgn}(k_{\parallel})& $k_{-}<|k_{\parallel}|<k_{+}$\\ \nonumber
2\text{sgn}(k_{\parallel})& $0<|k_{\parallel}|<k_{-} $.
\end{numcases}

Here, $k_{\pm} $ are the solutions of the equation $\text{det}[q(k_{\pm}, k_{\perp})]=0$ with $0<k_{-}< k_{+}$. When $|N_{l,BDI}(k_\parallel)|=1$, there is a single Majorana mode associated with a momentum quantum number $k_{\parallel}$ localized on the [110] edge. The regimes with $|N_{l,BDI}(k_\parallel)|=1$ are highlighted in red in Fig.2b. When $|N_{l,BDI}(k_\parallel)|=2$, there is a single fermionic mode localized on the edge. The flat band regime, which is not highlighted in red in Fig.2b, has $|N_{l,BDI}(k_\parallel)|=2$. It is important to note that when the Rashba terms are zero, $ k_{+}=k_{-} $. Therefore, $|N_{l,BDI}(k_{\parallel})|$ is always 0 or 2 and there can be no Majorana modes as shown in Fig.2a.

\section{D classification at finite magnetic field}
In this section, we show that the zero energy fermionic modes can be lifted to finite energy by an in-plane magnetic field. Meanwhile, the Majorana modes stay at zero energy even in the presence of an in-plane magnetic field which is parallel to the sample edge. The energy spectra in the presence of a magnetic field are shown in Fig.2c and Fig.2d. To understand the energy spectra, we note that the time-reversal like symmetry $T_{1d}$, which protects the zero energy fermionic modes with $|N_{l,BDI}(k_{\parallel})|=2$, is broken by a magnetic field. However, even though the $T_{1d}$ symmetry is broken, the particle-hole like symmetry $P_{1d}$ is preserved if the in-plane magnetic field is applied parallel to the [110] edge direction. In the presence of Rashba SOC, the 1D Hamiltonians $H_{l,k_{\parallel}}(k_\perp)$ are in the D class \cite{SRFL, Teo}. Then, $H_{l,k_{\parallel}}(k_\perp)$ can be classified by the $Z_{2}$ topological invariant \cite{Kitaev}$N_{l,D}(k_{\parallel})$ defined by
\begin{equation}
(-1)^{N_{l,D}(k_{\parallel})}=\text{sgn} [ \text{Pf}B_{l}(k_{\parallel},0)\text{Pf}B_{l}(k_{\parallel},\pi)]  \label{ND}
\end{equation}
where Pf denotes the Pfaffian and the matrix $ B_{l}(\bm k) $ is defined by $ B_{l}(k_{\parallel}, k_{\perp})=H_{l}( k_{\parallel}, k_{\perp}){P}_{1d}$, which is antisymmetric at $ k_{\perp}=0 $ and $ k_{\perp}=\pi$. It can be shown that $N_{l,D}(k_{\parallel})=1$ when $\text{Re}\lbrace\text{det}[q_{l}(k_{\parallel}, 0)]\rbrace-V_{\parallel}^2 <0 $. 

Consequently, there can be a finite range of $k_{\parallel}$ that $N_{l,D}(k_{\parallel})=1$ and there are Majorana modes associated with quantum numbers $k_{\parallel}$ localized on the [110] edge. The regimes with $N_{l,D}=1$ are highlighted in red in Fig.2d. It is evident from Fig.2d that the Majorana flat bands are robust to an in-plane field parallel to the [110] edge. 

In the absence of Rashba SOC, as the spin rotation symmetry along $y$-axis is preserved, the 1D Hamiltonians are in A class \cite{SRFL, Teo}, which is topologically trivial in 1D. As a result, there are no flat bands in the spectrum as shown in Fig.2c. 

It is important to note that this reduced symmetry from BDI to D class in the presence of a magnetic field is important for the determination of the stability of Majorana fermions in the multilayer cases.

\section{Tunnelling conductance}
It has been pointed out previously that Majorana fermions induce quantized ZBCPs in Andreev reflection type experiments of topological superconductors with a bulk gap \cite{Law, Wimmer}. Remarkably, ZBCPs  that are possibly due to Majorana fermion induced resonant Andreev reflections have been observed recently \cite{ Kou, Deng, Das, Marcus} in superconductor-semiconductor heterostructure \cite{LSD, Alicea2, ORV, Potter2}. Therefore, we expect that Majorana fermions localized on the edges of a nodal superconductor can also induce  almost resonant Andreev reflections as explained in Ref.\cite{Wong}. In the nodal $d_{x^2-y^2}$-wave superconductor case however, the appearance of Majorana edge modes are accompanied by the existence of zero energy fermionic states, which are localized at the same edge. These zero energy fermionic modes can cause ZBCPs as well \cite{Hu, Tanaka2}. Therefore, it is important to find a way to distinguish the Majorana edge modes from the fermionic edge modes.

Fortunately, as shown in the previous section, the zero energy fermionic edge modes of a usual $d_{x^2-y^2}$-wave superconductor can be lifted by an external magnetic field to finite energy. As a result, the ZBCPs caused by fermionic edge modes are split into two conductance peaks at finite bias as shown in Fig.3c. These finite bias conductance peaks have been observed experimentally \cite{Tanaka}. However, the Majorana edge modes, which appear in the presence of Rashba SOC, stay at zero energy even in the presence of an in-plane magnetic field parallel to the [110] edge. Therefore, we expect the ZBCPs to survive in the presence of a magnetic field. Moreover, since the zero energy fermionic modes are lifted to finite energy in the presence of a magnetic field, we expect the ZBCP that originated from the fermionic modes to split into two peaks at finite voltage bias. As a result, we expect a triple-peak structure for the tunnelling conductance in the presence of Majorana fermions. The tunnelling conductance in the presence of Majorana fermions is shown in Fig.3d. For comparison, the tunnelling conductance with zero magnetic field, in the absence and presence of the Rashba SOC, are shown in Fig.3a and Fig.3b respectively.

The results shown in Fig.3 are calculated using the lattice Green's function method \cite{Lee1, Lee2}. A schematic picture of the tunnelling experiment is shown in Fig.1a. The superconductor is described by a tight-binding lattice model of $H_{l,tb}$ in Eq.2 with $N_{\parallel}$ sites along the [110] edge and $N_{\perp}$ sites perpendicular to the [110] edge. A semi-infinite normal lead with width $N_{L}$, described by $H_{l,t}$ in Eq.2 with hopping amplitude $t_{lead}$, is attached to the [110] edge of the superconductor. The barrier between the normal lead and the superconductor is simulated by a reduced hopping strength $t_{h}$ between the neighbouring sites of the lead and the superconductor.

The differential conductance from the normal lead into the edge of the nodal superconductor can be written as:
\begin{equation}
\frac{dI}{dV}=\frac{e}{h}\frac{d}{dV}\int_{-\infty}^{\infty} {d\epsilon } T_{N}(\epsilon) (f_e  - f) + T_{A}(\epsilon) (f_e-f_h).
\end{equation}
Here, $V$ denotes the voltage difference between the lead and the superconductor.  $f(\epsilon)$ is the Fermi distribution function $f(\epsilon)=[\exp(\epsilon/k_{B}T)+1]^{-1}$ and $f_{e/h}(\epsilon)=f(\epsilon\mp eV)$. $T_{A}(\epsilon)$ denotes the Andreev reflection amplitude and $T_{N}(\epsilon)$ denotes the electron tunnelling amplitude from the normal lead to the superconductor. $T_{A}$ and $T_{N}$ are calculated using the standard recursive Green's function method \cite{Lee1, Lee2}. It is important to note that $T_{N}(\epsilon)$ is zero for a fully gapped superconductor when $\epsilon$ is smaller than the pairing gap. However, for a nodal superconductor, $T_{N}$ is not zero in general \cite{Wong}. 

\begin{figure}
\centering
\includegraphics[width=3.2in]{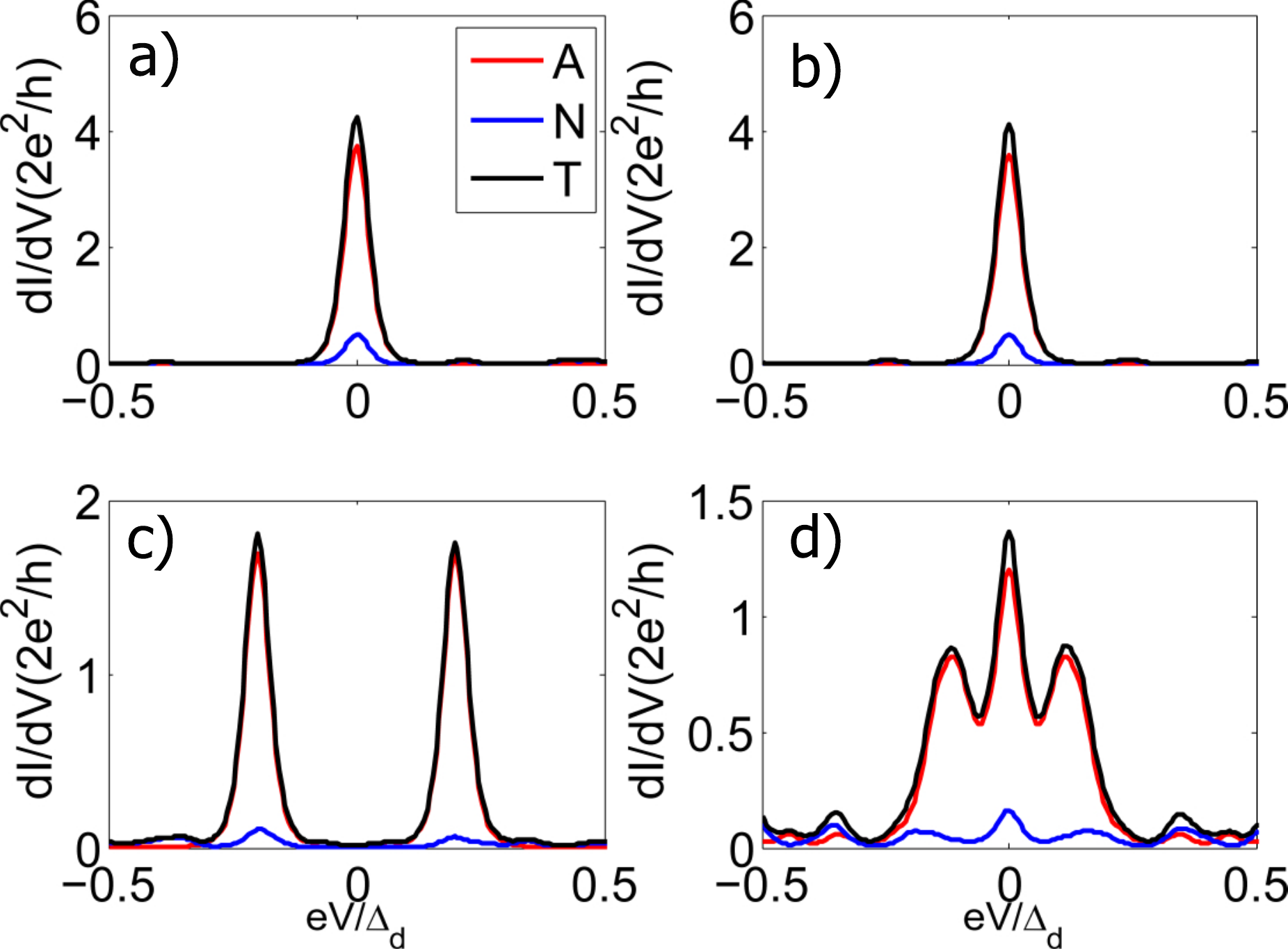}
\caption{\label{Fig3} The tunnelling conductance of $d_{x^2-y^2}$-wave superconductors as functions of the bias voltage. A and N denote the Andreev reflection and normal electron tunnelling contributions to the differential conductance respectively. T denotes the total differential conductance. For all the figures, $N_{\parallel}=60$, $N_{\perp}=150$, $N_{L}=30$. The parameters of the model are: $t=10$, $\Delta_{d} =2, t_{lead}=5t, t_{h}=0.38t$. The temperature is set to $ k_{B}T=0.015\Delta_{d} $. a) In the absence of Rashba SOC with $\alpha_{l}=0$ and ${\bm V}=\bm 0$. The ZBCP is induced by zero energy fermionic modes. b) In the presence of Rashba SOC with $\alpha_{l}=2.5\Delta_{d}$ and ${\bm V}=\bm 0$. The ZBCP is due to both the Majorana and fermionic zero energy states. c) $\alpha_{l}=0$, ${V}_{\parallel}=0.2\Delta_{d}$. The ZBCP is split into two conductance peaks at finite voltage. d)  $\alpha_{l}=2.5 \Delta_{d}$ and ${{ V}_{\parallel}}=0.2\Delta_{d}$. The ZBCP is due to Majorana fermions and the conductance peaks at finite voltage are due to finite energy fermionic edge states.}
\end{figure}

\section{Multilayer cases}
It is shown above that Rashba SOC is essential for the appearance of Majorana fermions in $d_{x^2-y^2}$-wave nodal superconductors. Unfortunately, a single layer $d_{x^2-y^2}$-wave superconductor with strong Rashba SOC is yet to be identified. On the other hand, superconducting thin films of $\text{CeCoIn}_{5}$, with $d_{x^2-y^2}$-wave pairing and strong SOC, were fabricated recently\cite{Matsuda}. It is important to note that bulk $\text{CeCoIn}_{5}$ has a crystal point group symmetry $D_{4h}$, which respects inversion symmetry, and Rashba type SOC cannot appear in the bulk. Nevertheless, as pointed out by Maruyama et al.\cite{Sigrist}, inversion symmetry is locally broken on the surface layers of $\text{CeCoIn}_{5}$ which give rise to Rashba SOC on the surface layers. In this case illustrated in Fig1b, the top and bottom layers of $\text{CeCoIn}_{5}$ can be described by $H_{l}$ in Eq.1 with different Rashba strength and the Rashba strength in the middle layers can be neglected. Motivated by these experimental and theoretical studies, we investigate two layers of $d_{x^2-y^2}$-wave superconductors with Rashba SOC which are coupled by electron hopping. Due to the fact that the local inversion symmetry of the top and bottom surface layers can be broken in different ways due to the difference between the substrate and the vacuum as shown in Fig.1b, we allow the Rashba strength on the two layers to be different in general. When this happens, the global inversion symmetry of the superconductor is broken.

The tight-binding Hamiltonian of the bilayer system is $ H_{\text{bilayer},tb}=H_{1,tb} + H_{2,tb}+H_c$, where $H_{l,tb}$ are described in Eq.2 and $H_{c}$ is the coupling between the two layers with:
\begin{equation}
H_c =-t_{c}\sum_{\bm R s}c_{1 \bm R s}^{\dagger}c_{2 \bm R s}+h.c. 
\end{equation}
After Fourier transform, we obtain the Hamiltonian in the momentum space $H_{\text{bilayer}}(\bm k)$. Before discussing the details of the calculations, we note that $H_{\text{bilayer}}(k_{\parallel}, k_{\perp})$ satisfies the time-reversal like symmetry and particle-hole like symmetry as before with $T_{1d}=-\sigma_{0}\otimes \tau _{z} \otimes I_{0} K$ and $P_{1d}=-\sigma_{y} \otimes \tau_{y} \otimes I_{0}K$. Here, $I_{0}$ is the identity matrix acting on the layer index. As a result, the Hamiltonian $H_{\text{bilayer}}(k_{\parallel}, k_{\perp}) =H_{\text{bilayer},k_{\parallel}}(k_{\perp})$, with $k_{\parallel}$  as a tuning parameter, is in the BDI class in the absence of an external magnetic field and in D class in the presence of an in-plane magnetic field parallel to the [110] edge.

As pointed out by Schnyder et al. \cite{SRFL}, BDI class Hamiltonians are classified by integers. Therefore, if both the top and the bottom layers are topologically non-trivial for a particular $k_{\parallel}$ before turning on the coupling, the bilayer system is expected to be topologically non-trivial even when the two layers are coupled. Therefore, we expect that the bilayer system support Majorana flat bands when at least one of the layers is topologically non-trivial. This is in sharp contrast to the case of time-reversal breaking topological superconductors whereby coupling two topologically non-trivial superconductors renders the system topologically trivial \cite{Kitaev, Potter}.

In the presence of an in-plane magnetic field parallel to the [110] edge, the system is in D class which is classified by a $Z_{2}$ number. As a result, the bilayer system supports Majorana fermions only when there are regions of $k_{\parallel}$ where only one layer is topologically non-trivial. This happens when the global inversion symmetry of the system is broken as we show below.

The above argument can be verified by calculating the topological invariant $N_{\text{bilayer}, BDI}(k_{\parallel})$ and $N_{\text{bilayer},D}(k_{\parallel})$ of the BDI class and D class Hamiltonian respectively. $N_{\text{bilayer}, BDI}(k_{\parallel})$ can be calculated using Eqs.\ref{NBDI1} and \ref{NBDI2} with $q_{l}$ replaced by $q_{\text{bilayer}}$ where 
\begin{equation}
\begin{aligned}
\det q_{\text{bilayer}}(k_{\parallel} ,k_{\perp}) =&[\det q_{1}(k_{\parallel} ,k_{\perp})-t_{c}^2][\det q_{2}(k_{\parallel} ,k_{\perp})-t_{c}^2]\\
&-t_{c}^2(\alpha_{1}+\alpha_{2})^2(\sin^2 k_{\parallel}+\sin^2 k_{\perp}).
\end{aligned}
\end{equation}
When the interlayer coupling satisfies $|t_{c}/t|<<1$, it can be shown that $N_{\text{bilayer}, BDI} = N_{1, BDI} + N_{2,BDI}$.

In the presence of a magnetic field, $N_{\text{bilayer},D}(k_{\parallel})$ can be found using Eq.\ref{ND} with $\text{Pf}B_{l}$ replaced by $\text{Pf}B_{\text{bilayer}}$ where 
\begin{align}\nonumber
\text{Pf}B_{\text{bilayer}}(k_{\parallel},k_{\perp})=&[\text{Pf}B_{1}(k_{\parallel},k_{\perp})-t_{c}^2][\text{Pf}B_{2}(k_{\parallel},k_{\perp})-t_{c}^2]\\
&-t_{c}^2(\alpha_{1}+\alpha_{2})^2(\sin^2 k_{\parallel}+\sin^2 k_{\perp}).
\end{align}
for $k_{\perp}=0,\pi$. It can be shown that for small interlayer coupling $t_c$, $N_{\text{bilayer},D}= N_{1,D}+N_{2,D} (\text{mod}2)$.

Due to the symmetry classification, in the case of weak interlayer coupling, we expect $N_{\text{multilayer}, BDI} = \sum_{l} N_{l, BDI}$ and $N_{\text{multilayer},D}= \sum_{l}N_{l,D} (\text{mod}2)$ for the multilayer cases.

\begin{figure}
\centering
\includegraphics[width=3.2in]{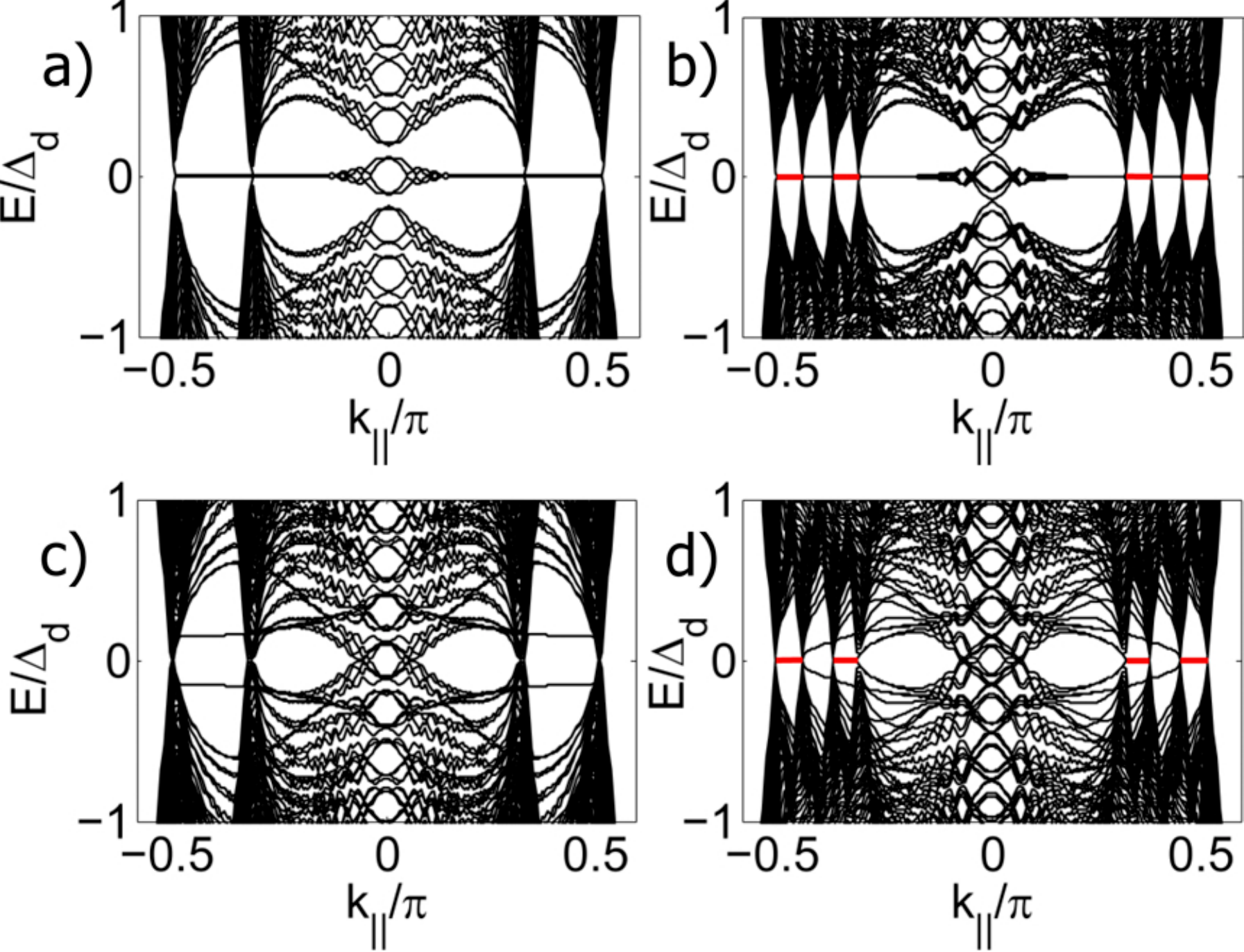}
\caption{\label{Fig4} The energy spectra for the bilayer case. $t=5 \Delta_{d}$, $\Delta_d=2$, $t_c=1.5\Delta_{d}$ for all the figures.  a) $\alpha_1 = -\alpha_2 =2.5\Delta_{d}$ and ${\bm V}=\bm 0$. The Majorana flat band regimes of the two layers overlap completely with each other. b) $\alpha_1= -5 \alpha_2=2.5\Delta_{d}$ and ${\bm V=\bm 0}$. The Majorana flat bands of the two layers overlap partially. c) Same parameters as in a) except ${V}_{\parallel}=0.3\Delta_{d}$. The Majorana and fermionic flat bands are lifted to finite energy. d) Same parameters as in c) except ${ V_{\parallel}=0.3 \Delta_{d}}$. The Majorana flat bands are lifted to finite energy for the range of $k_{\parallel}$ where both layers have Majorana flat bands. The stable Majorana flat band regime is highlighted in red.}
\end{figure}

The energy spectra of bilayer systems with periodic boundary conditions parallel to the [110] edge and open boundary conditions perpendicular to the edge are shown in Fig.4. Fig.4a shows the energy spectrum of the bilayer case with $\alpha_{1}=-\alpha_{2}$. In this case, the topological invariant for all $ k_{\parallel} $ satisfies $N_{1,BDI}=N_{2,BDI}$ resulting in the persistence of Majorana flat bands in the bilayer case. However, in the presence of a magnetic field, the symmetry is reduced from BDI class to D class and all the zero energy modes are lifted to finite energy as shown in Fig.4c. This is due to the fact that $N_{1,D}(k_{\parallel})=N_{2,D}(k_{\parallel})$ for all $k_{\parallel}$ and $N_{\text{bilayer},D }$ is always trivial for all $k_{\parallel}$.

If the global inversion symmetry is broken, the magnitude of the Rashba strengths are different $|\alpha_{1}| \neq |\alpha_{2}|$, there exist regimes in which $N_{1,D}(k_{\parallel}) \neq N_{2,D}(k_{\parallel})$. In this case, Majorana flat bands exist both in the absence and presence of a parallel magnetic field as shown in Fig.4b and Fig.4d respectively. Experimentally, if a thin film of CeCoIn$_{5}$ is grown on a substrate, as depicted in Fig.1b, the local inversion symmetry of the bottom layer and the top layer can be broken differently and result in $|\alpha_{1}| \neq |\alpha_{2}|$. Therefore, observing Majorana flat bands in the presence of an in-plane magnetic field is possible and a triple-peak structure of the tunnelling spectroscopy is expected, as in the single layer case.

\begin{figure}
\centering
\includegraphics[width=3.2in]{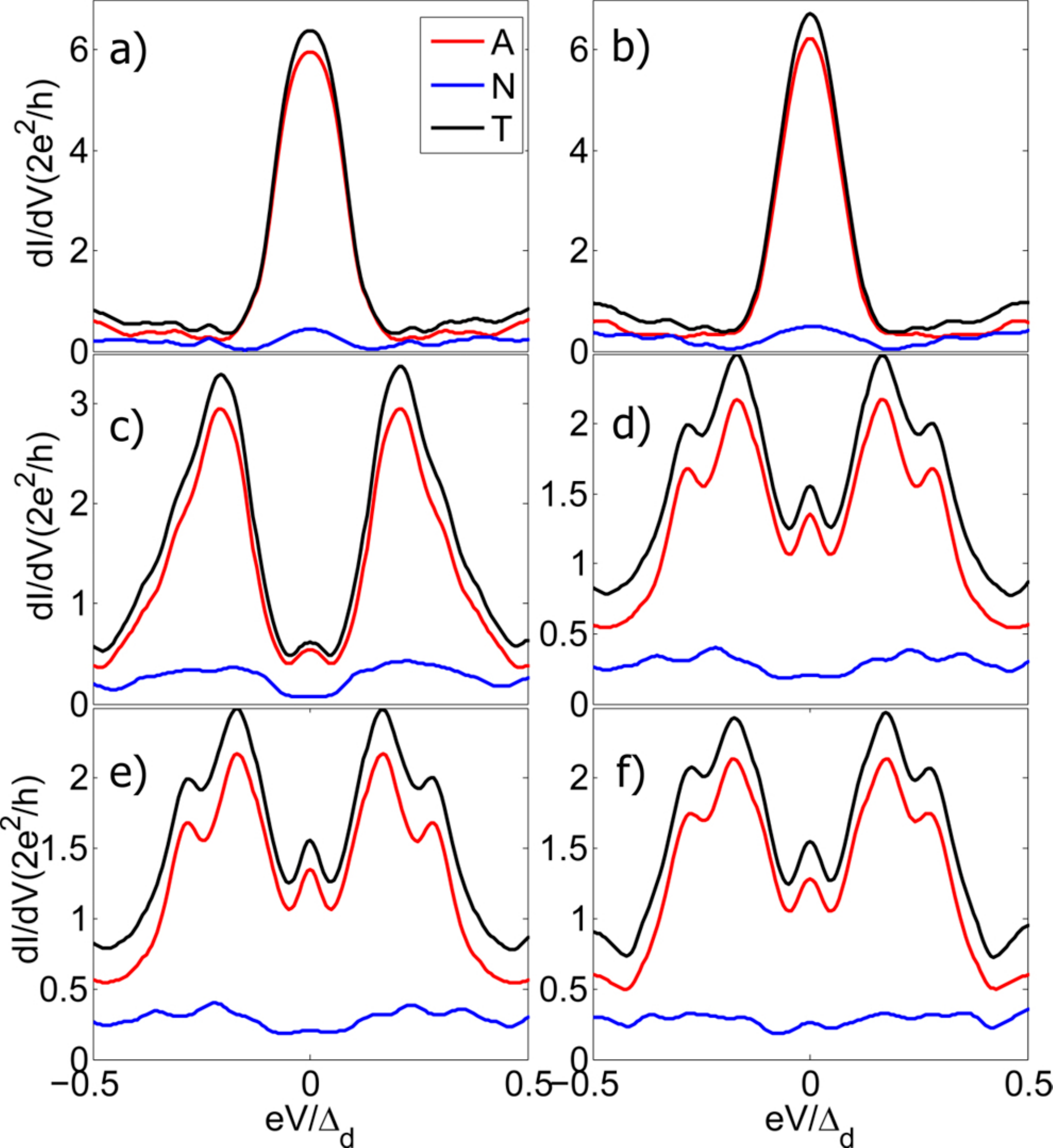}
\caption{\label{Fig5} The tunnelling conductance for the bilayer case. A and N denote the contributions of Andreev reflection and normal tunnelling to the differential conductance. T denotes the total differential conductance. For all the figures, $N_{\parallel}=100$, $N_{\perp}=150$, $N_{L}=40$, $t=5 \Delta_{d}$, $\Delta_d=2$, $t_c=1.5\Delta_{d}$,  $t_{lead}=5t$,  $t_{h}=0.38t$, $k_{B}T=0.015\Delta_d$. a) $\alpha_1 = -\alpha_2 =2.5\Delta_{d}$ and ${\bm V}=\bm 0$.  b) $\alpha_1= -5 \alpha_2=2.5\Delta_{d}$ and ${\bm V=\bm 0}$. c) Same parameters as in a) except ${V}_{\parallel}=0.3\Delta_{d}$. The ZBCP is split into two conductance peaks at finite voltage. d) Same parameters as in c) except ${ V_{\parallel}=0.3 \Delta_{d}}$.  e) Same parameters as in d) except that we have disorder with strength 0.5$\Delta_d$. f) Same parameters as in d) except that a $p$-wave pairing with strength 0.05$\Delta_d$ is added to the bilayer system.}
\end{figure}

The tunnelling conductance in the case with $|\alpha_{1}| = |\alpha_{2}|$ is shown in Fig.5a and Fig.5c in the absence and presence of a magnetic field respectively. As expected, the ZBCP is split into two peaks when $V_{\parallel} \neq 0$. The tunnelling conductance for $|\alpha_{1}| \neq |\alpha_{2}|$ is shown in Fig.5b and Fig.5d. Without magnetic field, there is a ZBCP due to the Majorana fermions and the usual zero energy fermions. With an in-plane magnetic field, the single ZBCP is split into three peaks with the central peak induced by Majorana fermions. These special magnetic field dependence of the ZBCPs can be used to probe Majorana fermions in nodal $d_{x^2-y^2}$ superconductors.

\section{Discussion and Conclusion}
It is important to note that both the fermionic flat bands and the Majorana flat bands are protected by translation symmetry. In the presence of disorder, the Majorana as well as the fermionic flat bands are broadened around zero energy. However, as shown in Fig.5e, the triple-peak structure in the finite magnetic field can still be observed as long as the Zeeman energy splitting is larger than the width of the disorder and thermal broadened tunnelling peak. Due to inversion symmetry breaking, it is possible that a $p$-wave pairing term can emerge \cite{Gorkov}. By symmetry analysis, such a $p$-wave pairing breaks the crucial mirror symmetry which protects the Majorana flat bands and thus the ZBCP caused by Majorana fermions will be further split into two peaks. The further splitting of the ZBCP can be regarded as a piece of evidence of the emergence of Rashba $p$-wave superconductor in the system.  However, if this splitting is smaller than the thermal broadening width, the two peaks will appear as one ZBCP at finite temperature. This case is shown in Fig.5f. Similarly, the symmetry protecting the Majorana flat bands is broken by magnetic fields with components perpendicular to the [110] edge. As a result, the Majorana flat bands will be lifted to finite energy and the ZBCP will be split into two peaks.

In conclusion, we pointed out that Majorana flat bands appear on the [110] edges in single or multi-layer $d_{x^2-y^2}$ superconductors with Rashba-spin orbit coupling. As long as global inversion symmetry is broken, the Majorana fermions can survive even in the presence of an in-plane magnetic field parallel to the [110] edge. The fermionic and Majorana edge states induce multiple peaks in tunnelling spectroscopy experiments with a ZBCP induced by Majorana fermions. Therefore, we propose that tunnelling experiments can be used to probe Majorana fermions in nodal $d_{x^2-y^2}$-wave superconductors.

\section{Acknowledgement}
The authors thank P. A. Lee, M. Sato and Y. Tanaka for inspiring discussions. This work is supported by HKRGC through Grant 605512 and HKUST SSCI Computational Science Initiative.

$^\dagger$ Email address: phlaw@ust.hk.

\end{document}